\documentclass[aps,prl,twocolumn,preprintnumbers,amsmath,amssymb,superscriptaddress]{revtex4-1}
\usepackage{graphicx}
\usepackage{dcolumn}
\usepackage{bm}

\newcommand\beq{\begin{equation}}
\newcommand\eeq{\end{equation}}
\newcommand\bseq{\begin{subequations}}
\newcommand\eseq{\end{subequations}}
\newcommand\bfig{\begin{figure}}
\newcommand\efig{\end{figure}}

\graphicspath{{pict/}{}}

\newcommand\pictc[5]{\begin{figure}
                   \centerline{
                   \includegraphics[width=#1\columnwidth,height=0.8\textheight,keepaspectratio]{#3}}
               \protect\caption{\protect\label{fig:#4} #5}
                \end{figure}            }

\newcommand\pict[4][1]{\pictc{#1}{!tb}{#2}{#3}{#4}}

\newcommand\rpict[1]{\ref{fig:#1}}

\newcommand\leqt[1]{\protect\label{eq:#1}}
\newcommand\reqtn[1]{\ref{eq:#1}}
\newcommand\reqt[1]{(\reqtn{#1})}

\begin{document}

\title{Compact surface Fano states embedded in the continuum of waveguide arrays}

\author{Steffen Weimann}
\affiliation{Institute of Applied Physics, Abbe Center of Photonics, Friedrich-Schiller-Universit\"{a}t Jena, Max-Wien-Platz 1, 07743 Jena, Germany}

\author{Yi Xu}
\affiliation{Nonlinear Physics Centre, Research School of Physics and
Engineering, Australian National University, Canberra ACT 0200, Australia}

\author{Robert Keil}
\affiliation{Institute of Applied Physics, Abbe Center of Photonics, Friedrich-Schiller-Universit\"{a}t Jena, Max-Wien-Platz 1, 07743 Jena, Germany}

\author{Andrey E. Miroshnichenko}
\affiliation{Nonlinear Physics Centre, Research School of Physics and
Engineering, Australian National University, Canberra ACT 0200, Australia}

\author{Stefan Nolte}
\affiliation{Institute of Applied Physics, Abbe Center of Photonics, Friedrich-Schiller-Universit\"{a}t Jena, Max-Wien-Platz 1, 07743 Jena, Germany}

\author{Andrey A. Sukhorukov}
\affiliation{Nonlinear Physics Centre, Research School of Physics and
Engineering, Australian National University, Canberra ACT 0200, Australia}

\author{Alexander Szameit}
\affiliation{Institute of Applied Physics, Abbe Center of Photonics, Friedrich-Schiller-Universit\"{a}t Jena, Max-Wien-Platz 1, 07743 Jena, Germany}

\author{Yuri S. Kivshar}
\affiliation{Nonlinear Physics Centre, Research School of Physics and
Engineering, Australian National University, Canberra ACT 0200, Australia}


\maketitle

\textbf{Wave localization at surfaces is a fundamental physical phenomenon with many important examples such as surface plasmon polaritons
~\cite{Boardman:Book:1982, Maier:Book:2007}, Dyakonov surface modes~\cite{Dyakonov:JETP:1988}, and Tamm states~\cite{Tamm:ZP:1932}. Although being of
different origin, these modes share the same property:  they decay exponentially away from the surface, with the eigenfrequencies of the surface
states being outside the band of the spatially extended propagating states.  Here we describe theoretically and observe experimentally a novel type
of surface states in a semi-infinite waveguide array with simultaneous Fano and Fabry-Perot resonances. We demonstrate that the Fano surface mode is
compact, with all energy concentrated in a few waveguides at the edge and no field penetration beyond the side-coupled waveguide. We show that by
broadening the spectral band in the rest of the waveguide array it is possible to suppress exponentially localized modes, whereas the Fano state
having the eigenvalue embedded in the continuum is preserved.}

Remarkably, in their seminal paper von Neumann and Wigner~\cite{Neumann:ZP:1929} explicitly constructed a potential that supports so-called ``bound
states in the continuum'' (BIC), i.e., a particular localized mode with the eigenfrequency inside the band of extended states. Since then, the
peculiar concept of BIC attracted a lot of attention in various branches of physics~\cite{Stillinger:PRA:1975,Friedrich:PRA:1985}, reaching a climax
with the direct observation of an electronic bound state above a potential well due to Bragg reflection in semiconductor
heterostructures~\cite{Capasso:Nature:1992}. In the optical domain it was shown that BIC can be generated in photonic crystals and optical waveguide
arrays~\cite{Longhi:EPJB:2007, Marinica:PRL:2008, Bulgakov:PRB:2008, Dreisow:PRL:2008, Longhi:JMO:2009, Bulgakov:PRB:2009, Otey:2009-231109:APL,
Bulgakov:PRB:2010}. The first true observation of a BIC in any system was carried out in an optical system: a planar optical waveguide array with two
side-coupled defects, in which an in-band defect state is formed by virtue of symmetry \cite{Plotnik:PRL:2011}. Recently, a special design of a
one-dimensional (1D) waveguide array with modulated inter-site coupling was suggested in order to obtain ``surface BIC''~\cite{mario:PRL:2012}.

The phenomenon of Fano resonance~\cite{Fano:PR:1961} provides a simple and universal means for obtaining bound states in the continuum, which can be
realized in optics through the addition of side-coupled waveguides or cavities~\cite{Miroshnichenko:2010-2257:RMP}. Indeed, at Fano resonance there
appears an effectively infinite potential barrier such that an incident wave exhibits complete reflection, and such mechanism is responsible for
light trapping studied in photonic crystal waveguides between two side coupled defects~\cite{Pan:APL:2008, Pan:PRB:2010, Pan:APL:2010}, at a cavity
placed near the waveguide termination~\cite{Noda:NatMater:2007}, and predicted for waveguide arrays with side-coupled
waveguides~\cite{Longhi:EPJB:2007}. In this regime the mode can be trapped even when its eigenvalue is embedded in a continuum spectrum of the optical
waveguide.

In this Letter, we present the theoretical analysis and the experimental observation of the formation of surface Fano states. To this end, we employ
an optical system: an array of evanescently coupled waveguides with a side coupled waveguide close to the lattice edge. We show that the Fano modes
are ``compact'', i.e., there is no wave penetration beyond the location of the side-coupled waveguide, in agreement with the principles of Fano
resonance based on complete destructive interference in transmission. Based on this property, we demonstrate that usual exponentially localized
defect modes can be suppressed by increasing the waveguide coupling past the side-coupled waveguide: These modes disappear when their eigenvalue
enters the broadened spectral band, in contrast to the Fano mode that remains trapped. Such a selective manipulation of modes sensitive to their
localization mechanism may find applications in various physical contexts, since Fano resonance is a general wave phenomenon.
\pict{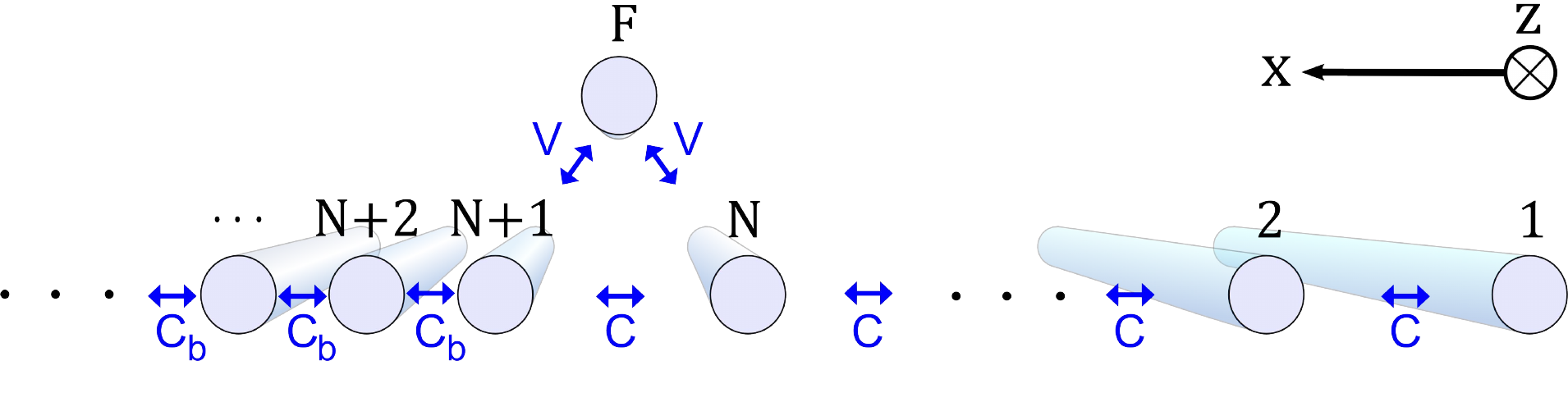}{figure1}{ (Color online) Sketch of a semi-infinite waveguide array, where a side-coupled waveguide is introduced to provide the Fano
resonance. }
We study the formation of Fano surface localized modes in a planar optical waveguide array with one side-coupled waveguide as shown in
Fig.~\rpict{figure1}. In analogy to the atomic physics context where Fano results were originally derived, this side-coupled waveguide is called the
``autoionizing" waveguide (AW). The waveguide structure consists of a 1D chain of identical waveguides. The propagation direction and the transverse
direction are $z$ and $x$, respectively. The chain is coupled to the AW with strength $V$. The part containing the waveguides $1$ to $N$ is a
resonator region. Here, the coupling coefficient between neighboring waveguides is $C$. We consider the general situation when the coupling strength
$C_{b}$ in the ``background" region, i.e., the semi-infinite array section containing the waveguides $N+1$, $N+2$, $\ldots$, is different. As we demonstrate below, the variation of the background coupling enables selective manipulation of Fano modes and regular exponentially localized modes.

We model the evolution of the optical mode amplitudes along the propagation direction using a coupled-mode approach with nearest-neighbor
coupling~\cite{Miroshnichenko:PRE:2005},
\begin{equation} \leqt{eq1}
     \begin{split}
        &i\frac{\partial\psi_j}{\partial z} + C(\psi_{j+1}+\psi_{j-1}) = 0, \quad 1 \le j < N\\
        &i\frac{\partial\psi_F}{\partial z} + V(\psi_{N+1}+\psi_{N}) = 0, \\
        &i\frac{\partial\psi_N}{\partial z} + V\psi_F+C(\psi_{N-1}+\psi_{N+1}) = 0, \\
        &i\frac{\partial\psi_{N+1}}{\partial z} + V\psi_F+C(\psi_{N}+\psi_{N+2}) = 0, \\
        &i\frac{\partial\psi_j}{\partial z} + C_b(\psi_{j+1}+\psi_{j-1}) = 0, \quad j > N+1 .
     \end{split}
\end{equation}
where $\psi_j$ describes the optical mode amplitude in the $j$-th waveguide. We put $\psi_0 \equiv 0$ due to the array termination. In this
formulation, the spectrum is centered around the propagation constant of an isolated waveguide.

We first consider the case of a homogeneous coupling in the waveguide array, $C_b=C$. Then, the eigenvalues $\beta$ in the semi-infinite array for a
continuous band in the region $\beta\in[-2C,2C]$. By virtue of the dispersion relation $\beta/C=2\cos(\kappa)$, each supermode can be
constructed from plane waves $\exp(i\kappa n)$ with wavenumber $\kappa$. The waveguide $F$ is a defect to the 1D chain and plane waves incident on
the triangle of waveguides exhibit a transmission which is dependent on the value of $\kappa$. It is convenient to represent the field inside the
resonator region ($n=1,\ldots,N$) in terms of incident and reflected waves as $\psi_n= \exp(i\beta z) [ \exp( i (n-N)\kappa) + R \exp( - i(n-N)
\kappa)]$ and in the background region ($n=N+1,N+2,...$) the field will be the transmitted wave $\psi_n = \exp(i\beta z) T \exp[ i (n-N-1) \kappa]$.
One finds from Eq.~\reqt{eq1} with $C_b=C$ that the transmission and reflection coefficients are
\begin{equation} \leqt{TR}
 \begin{split}
    T &= i e^{i \kappa} \left[ C^2 \sin(3 \kappa /2) + (V^2-C^2) \sin(\kappa/2) \right] D^{-1}, \\
    R &= e^{i \kappa} \cos(\kappa/2) V^2 D^{-1} ,
 \end{split}
\end{equation}
where $D = C^2 [ i \sin(3 \kappa /2) + \cos(\kappa/2)] + (V^2-C^2) \exp(i \kappa / 2)$. We note that the transmission vanishes at certain value of
$\kappa = \kappa_F$, which corresponds to a Fano resonance~\cite{Miroshnichenko:PRE:2005}. At resonance, we have $R(\kappa_F) = \exp(i \kappa_F)$,
i.e. $\kappa_F$ defines the phase of the reflection coefficient. In order to use the Fano resonance to trap light in the resonator region, it is
necessary to also satisfy the Fabry-Perot condition for one round-trip of the plane wave between the surface of the chain and the AW, which is
formulated as
\begin{equation} \leqt{FP}
   2 (N-1) \kappa_F + \kappa_F + 2 \kappa_F = \kappa_F (2 N + 1) = 2 \pi m,
\end{equation}
where $m$ is an integer. In this expression, the first term corresponds to phase accumulation for the waves propagation between the waveguides
$n=1,\ldots,N$, the second term is the reflection phase from the Fano defect, and the third term is the reflection phase at the boundary $n=1$. We
find that the Fano and Fabry-Perot resonance conditions can be satisfied simultaneously when $V = C$. In this
case $k_F = 2 \pi /3$, and the Fabry-Perot resonances occur for $m=1,3,9,\ldots$, corresponding to the number of waveguides in the resonator region
$N = 1 + (3/2) (m-1) = 1,4,7,\ldots$

In the following, the case $N=4$ is studied numerically and experimentally. For the experimental study, the waveguide array has to be finite, and we
consider an array with all in all $J=49$ waveguides ($48$ waveguides in the 1D chain plus the AW). We numerically calculate the spectrum of
eigenmodes, and present it in Fig.~\rpict{figure2}(a). We find extended eigenstates with eigenvalues $\beta\in[-2C,2C]$ representing a
quasi-continuum. Furthermore, as indicated by arrows in the figure we identify two localized states, which arise from the coupling of the AW with the
1D chain. The state indicated by a red arrow is a defect state residing inside the gap above the band of eigenvalues and it is thus exponentially
localized ($\beta=2.38C$). The second state, indicated by the blue arrow, occurs at the eigenvalue corresponding to the Fano resonance, at it has a
fundamentally different field structure with the zero amplitude in the background, i.e., this state strictly exists only between the edge and the AW.
This property is referred to as ``compactness''. Furthermore, the propagation constant of this Fano-compact state (FCS) resides inside the allowed
propagation band. The Fano resonance does not break up the band of eigenvalues of the unperturbed 1D chain which makes the FCS a bound state in the
quasi-continuum of the 1D chain.
\pict{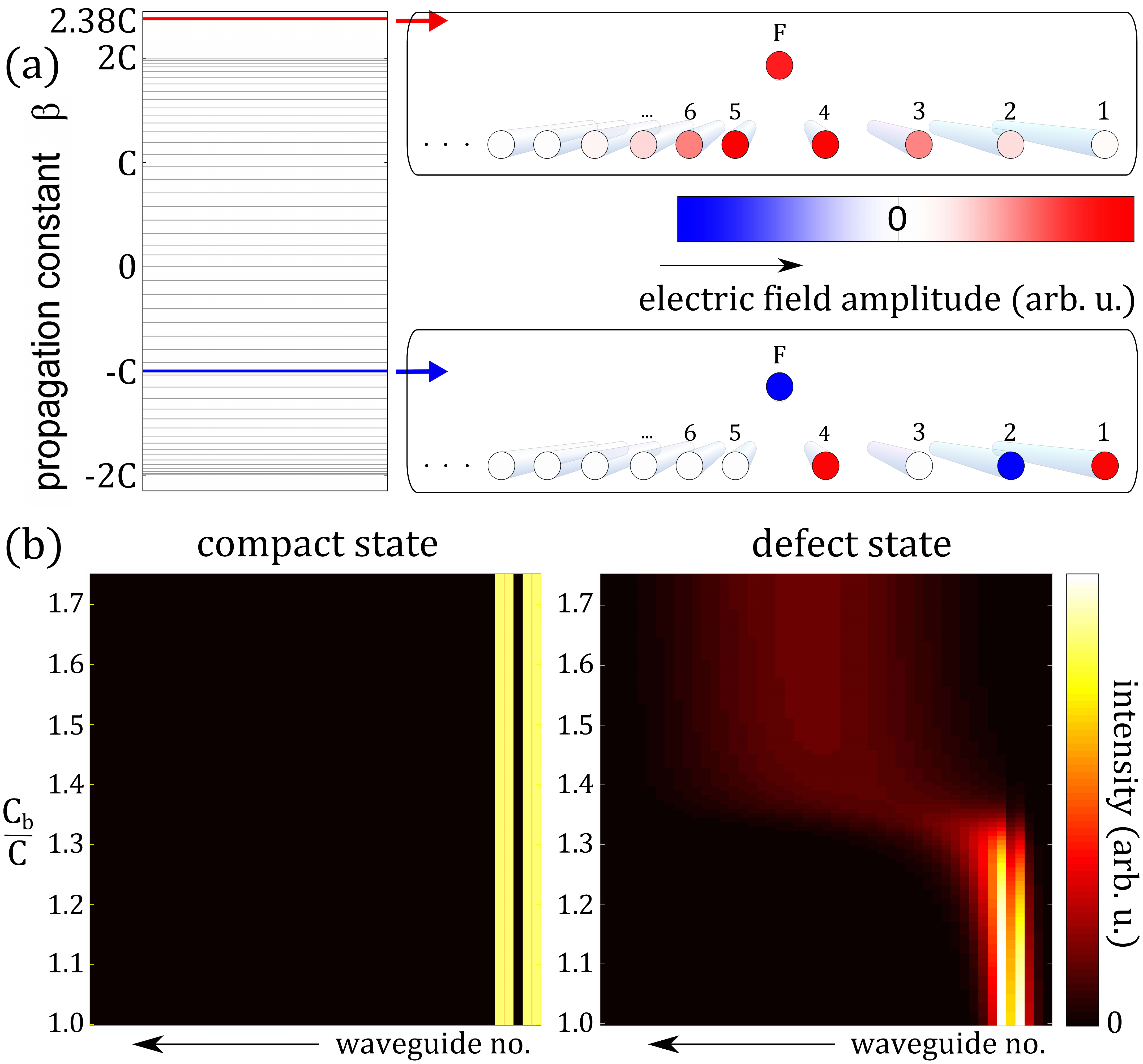}{figure2}{ (Color online) (a)~Eigenvalue spectrum of a structure with $49$ homogeneously coupled waveguides ($C_b=V=C$). The amplitude
profiles of the exponentially localized defect state (upper row) and the FCS (lower row) are shown on the right. (b,c)~The profile of (b)~FCA and
(c)~defect state as functions of $C_b$. The limit where the exponentially localized state ceases to exist is $C_b^{cr}=1.36C$. For $C_b\geq C_b^{cr}$
the defect state is absorbed in the quasi-continuum and hence spreads over the entire array. }
The vanishing of the electric field in waveguides $N=5,6,7,\ldots$ does not allow any interaction of the FCS with the residual lattice. Therefore the
state is completely insensitive to the lattice structure in the background. Hence, the FCS does not change if $C_b$ or the number of waveguides in
the background is varied. Figure~\rpict{figure2}(b) demonstrates the change of the FCS (left) and the defect state (right) with increasing $C_{b}$.
By an increase of $C_b\geq C$ the propagation band broadens until the exponentially localized state is absorbed into the band, whereas the FCS
remains unchanged at $\epsilon_{F}=-1$. The critical ratio of the coupling constants for which the exponentially localized state ceases to exist is
$C_b^{cr}/C=1.36$.

A light beam coupled to the waveguide array will excite a superposition of different modes. The amplitudes of eigenstates will be proportional to the
overlap of their spatial profiles with the input field distribution. For a single-waveguide excitation of the first site in the above structure, the
amount of intensity dedicated to the FCS is $25\%$. The defect state is only excited with a fraction of $0.37\%$ of the input intensity when $C_b=C$,
and this decreases further when $C_b>C$.

For our experiments, we fabricated several waveguiding structures in fused silica using the femtosecond laser direct-writing approach
\cite{Szameit:DiscreteOptics}. The samples are $10\text{cm}$ long, which is the maximum length feasible with our current fabrication technology. The
experimental investigations now deal with two separate issues. First, observing the Fano resonance in an array with homogeneous coupling in the
entire structure. Second, proving the compactness of the FCS by varying the coupling constant $C_b$ in the background. In order to directly observe
the light evolution inside the structures we employ a fluorescence microscopy technique \cite{Szameit:QuasiIncoherence}. Among the possible
single-waveguide excitations, the excitation of the first waveguide at the edge of the structure has the largest overlap with the FCS compared to the
overlap with the defect state. In our experiments light at $\lambda=633\text{nm}$ is launched into this waveguide using fiber butt coupling.
\pict{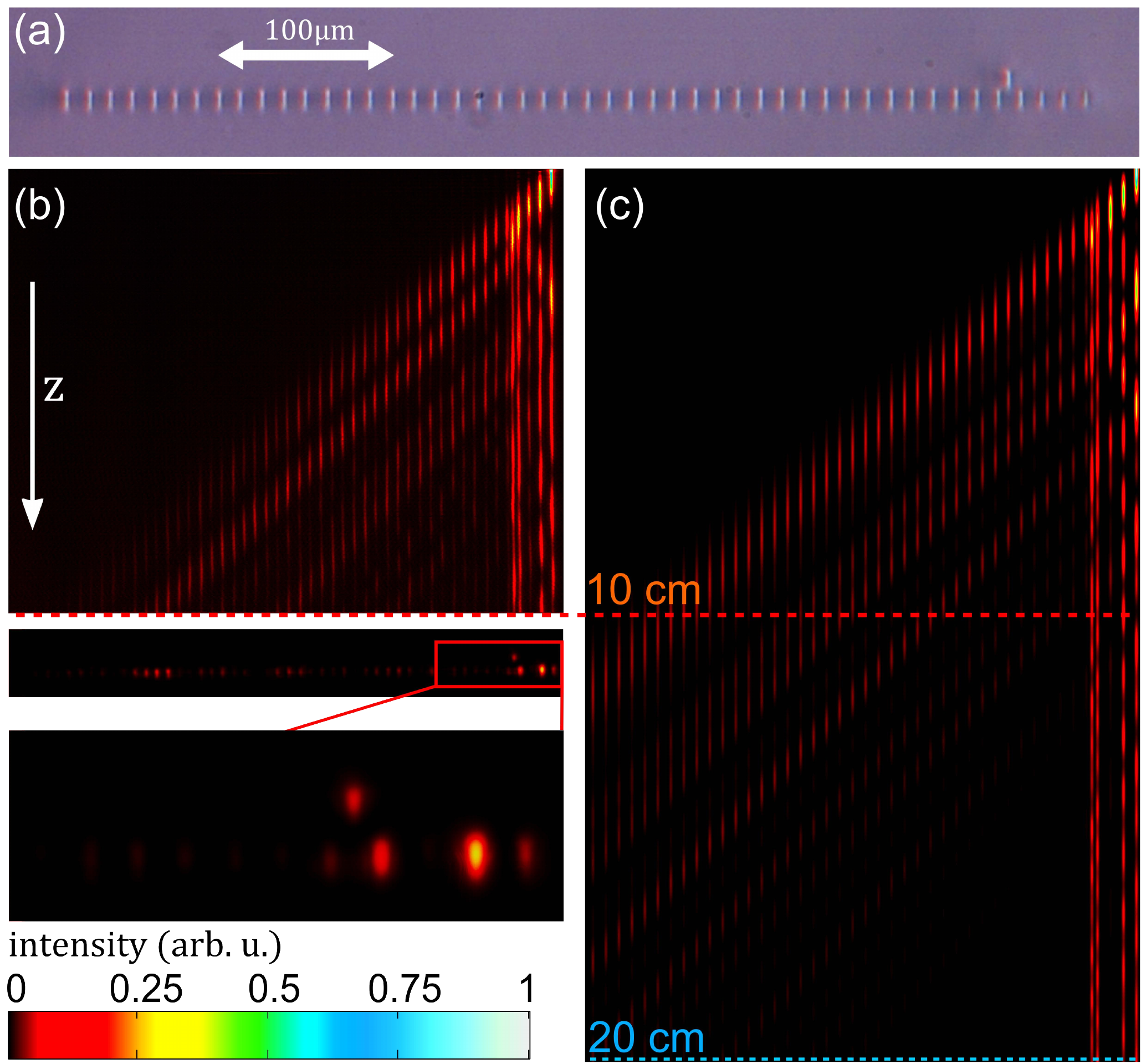}{figure4}{ (Color online) (a) Microscope image of the front facet of the processed fused silica sample serving as the experimental
realization of the structure. (b) Corresponding fluorescence microscopy measurement and the near-field image of the output facet. (c) Computed
intensity evolution for $20$cm of propagation. The fluorescence images are normalized to their respective peak value. }
In a first set of experiments the spacing between neighboring waveguides of the chain is fixed to $13\mu$m, such that $C_b=C$. This results in a
coupling of about $C=0.2\text{mm}^{-1}$. Figure~\rpict{figure4}(a) shows the microscope image of the front facet of the fabricated array. Due to the
strongly elliptical shape of the waveguides, resulting from the fabrication process, the coupling strength depends not only on the separation of the
guides but also on their orientation with respect to the coupling direction \cite{Szameit:DirectionalCoupling}. The vertical offset ($h$) of the AW
with respect to the chain was carefully tuned to match the condition $V=C$. Equal coupling was achieved for $h=({\sqrt{3}}d/2+2.6)\mu\text{m}$.

The experimental data is shown in Fig.~\rpict{figure4}(b). When launching light into the edge waveguide of the structure, several eigenstates of the
system are excited, including the FCS. Due to the reflection at the defect caused by the AW, the light carried by the extended eigenstates is
transmitted gradually into the background. After $10$cm propagation the intensity pattern in the resonator region is not perfectly reduced to the
intensity pattern of the FCS. The transient behavior within the first five waveguides still includes intensity oscillations in each waveguide.
However, the output pattern already exhibits the characteristic ``dark'' third waveguide [Fig.~\rpict{figure4}(b), lower part] identifying the FCS.
The simulations, shown in Fig.~\rpict{figure4}(c), confirm the measured behavior. Even after 20cm propagation, the FCS is not established perfectly
but the data in Fig.~\rpict{figure4} is sufficient to confirm the existence of the FCS.
\pict{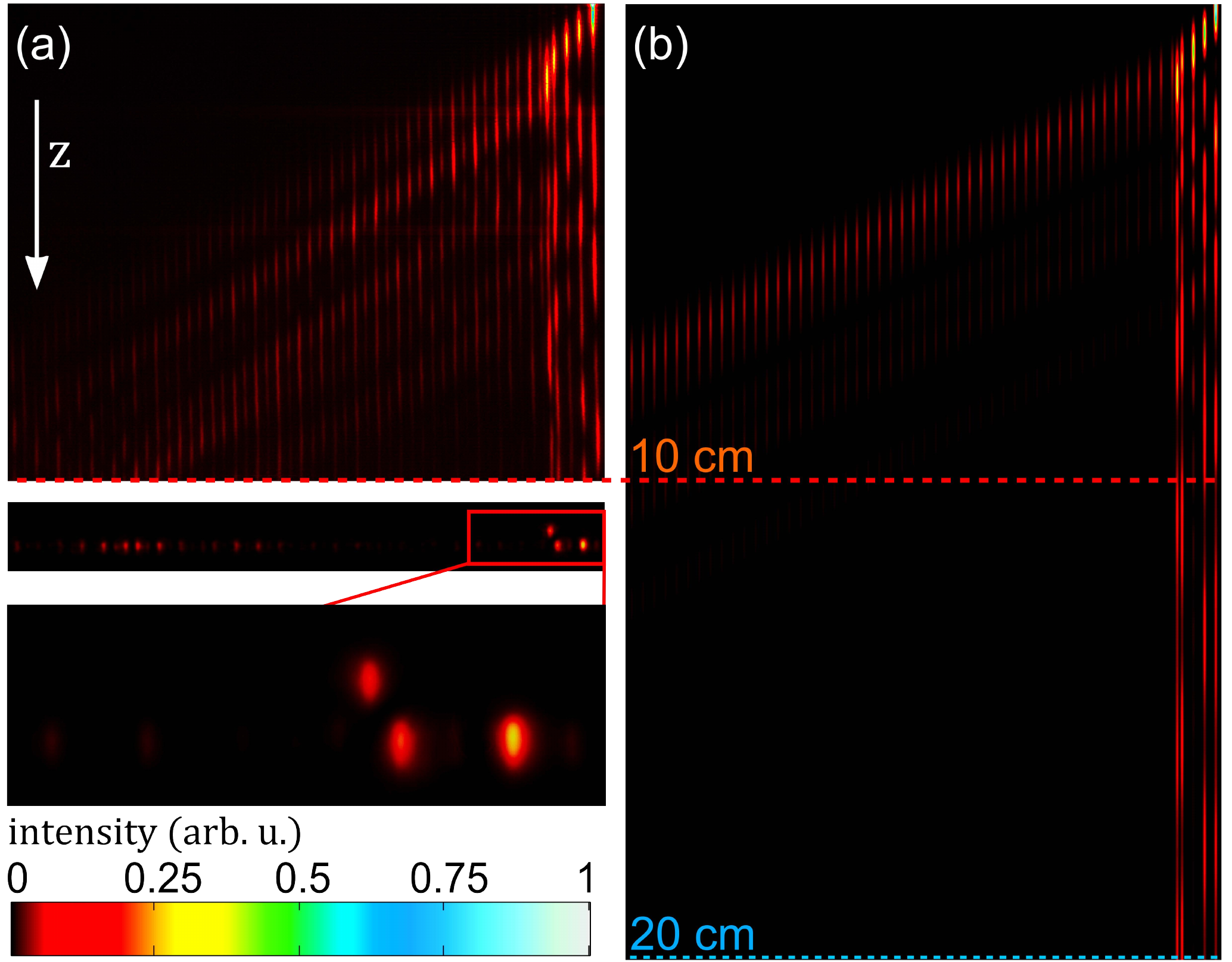}{figure5}{ (Color online) Light evolution in a structure, where the coupling in the background region is increased to $C_{b}/C=1.75$. (a)
Fluorescence microscopy measurement and the near-field image of the output facet. (b) Computed intensity evolution for $20$cm of propagation. Only
the fluorescence images are normalized to their respective peak value. }
In the next experiment, we aim to demonstrate the ``compactness'' of the FCS. To this end, in a second sample we reduced the spacing in the
background region to $11\mu\rm{m}$ which corresponds to $C_b\approx 0.35\text{mm}^{-1}$, whereas in the region close to the surface $13\mu$m spacing
is reproduced. Therefore, the Fano resonance is expected at a similar height $h$ of the AW, compared to the first structure. Indeed, $V=C$ was here
achieved for $h=({\sqrt{3}}d/2+2.2)\mu\rm{m}$. The experimental data in Fig.~\rpict{figure5}(a) shows that an intensity distribution in the first
five waveguides evolves to a final state which is very similar to that observed in the first experiment, c.f. Fig.~\rpict{figure4}(a). Our
simulations of the light evolution in this structure [Fig.~\rpict{figure5}(b)] confirm this trend. This is the experimental confirmation that the FCS
is independent of the topology of the background region of the 1D waveguide chain and, thus, the FCS is compact. Nevertheless, the transient behavior
changes slightly. This is reasonable since the occupation of the eigenmodes has changed and except for the FCS, the profiles of eigenmodes are
sensitive to the value of $C_b$.

In this work, we have theoretically predicted and experimentally demonstrated the existence of a compact bound state in the continuum of
one-dimensional waveguide array. This state is confined to a finite region of the array and exhibits no exponentially decaying tails. We found the
condition when all localized states other then the compact state cease to exist.

\textbf{Acknowledgments}\\ This work was support by the Australian Research Council through Future Fellowship program (including FT100100160) and the
German Ministry of Education and Research (Center for Innovation Competence programme, grant 03Z1HN31). R.~Keil was supported by the Abbe School of
Photonics.

\textbf{Methods}\\
Our samples are fabricated in bulk fused silica wafers using the femtosecond laser direct-write approach
\citep{Szameit:DiscreteOptics} employing a Ti:Sapphire Mira/RegA laser system (Coherent Inc.) operating at a wavelength of 800 nm, a repetition rate
of 100~kHz and a pulse length of 170~fs. The light is focused inside the sample by a $20\times$ microscope objective (NA=0.35), and by continuously
moving the sample using a high-precision positioning system (Aerotech Inc.) the waveguides are created by the induced refractive index increase. For
the fabrication of our samples, the pulse energy was adjusted to 320~nJ and the writing velocity was set to 1.5 mm/s.

\textbf{Author contributions}\\
S. W. fabricated the samples and performed the experiments. S. W. and A. S. analyzed the data. Y. X., A. E. M., and A. A. S. developed the theory.
All authors co-wrote the manuscript.

\textbf{Additional information}\\
Correspondence and requests for materials should be addressed to A.S. (alexander.szameit@uni-jena.de)

\textbf{Competing financial interests}\\
The authors declare no competing financial interests.

\end{document}